%% file: main.tex
\documentclass[10pt,journal,compsoc]{IEEEtran}
\usepackage{balance}
\usepackage{listings}
\usepackage{booktabs} % For formal tables
\usepackage{color-edits}
\usepackage{caption}
\usepackage{bm}

\makeatletter
\def\lst@makecaption{%
  \def\@captype{table}%
  \@makecaption
}
\makeatother

\ifCLASSOPTIONcompsoc
  \usepackage[nocompress]{cite}
\else
  \usepackage{cite}
\fi

\ifCLASSINFOpdf
\usepackage[pdftex]{graphicx}
\graphicspath{{img/}}
\else
\usepackage[dvips]{graphicx}
\fi

\usepackage{url}
\usepackage{multirow}

\usepackage{amssymb}
\usepackage{amsmath}

\usepackage{color-edits}

\addauthor{pg}{red}
\addauthor{ef}{blue}

\usepackage{algorithmic}

\usepackage{array}
\ifCLASSOPTIONcompsoc
 \usepackage[caption=false,font=footnotesize,labelfont=sf,textfont=sf]{subfig}
\else
 \usepackage[caption=false,font=footnotesize]{subfig}
\fi

\newcommand{\vct}[1]{\boldsymbol{#1}} % vector

\usepackage{algorithmic}    	% use algorithmic
\usepackage[vlined,ruled]{algorithm2e}	% use algorithm2e
\usepackage{xcolor}

\SetAlFnt{\sffamily}

\SetAlCapFnt{\normalfont\sffamily\large}

\makeatletter
\renewcommand{\algocf@Vline}[1]{%     no vskip in between boxes but a strut to separate them, 
  \strut\par\nointerlineskip% then interblock space stay the same whatever is inside it
  \algocf@push{\skiprule}%        move to the right before the vertical rule
  \hbox{\bgroup\color{cyan}\vrule\egroup%
    \vtop{\algocf@push{\skiptext}%move the right after the rule
      \vtop{\algocf@addskiptotal #1}\bgroup\color{cyan}\Hlne\egroup}}\vskip\skiphlne% inside the block
  \algocf@pop{\skiprule}%\algocf@subskiptotal% restore indentation
  \nointerlineskip}% no vskip after
\renewcommand{\algocf@Vsline}[1]{%    no vskip in between boxes but a strut to separate them, 
  \strut\par\nointerlineskip% then interblock space stay the same whatever is inside it
  \algocf@bblockcode%
  \algocf@push{\skiprule}%        move to the right before the vertical rule
  \hbox{\bgroup\color{cyan}\vrule\egroup%               the vertical rule
    \vtop{\algocf@push{\skiptext}%move the right after the rule
      \vtop{\algocf@addskiptotal #1}}}% inside the block
  \algocf@pop{\skiprule}% restore indentation
  \algocf@eblockcode%
}

\SetKwProg{Fn}{Function}{}{}
\addauthor{pg}{red}
\addauthor{nt}{blue}
\addauthor{kl}{green}
\addauthor{ef}{purple}

\begin{document}
\title{Capturing Edge Attributes via Network Embedding}

\author{Palash~Goyal, %, ~\IEEEmembership{Member,~IEEE,}
		Homa Hosseinmardi,
        Emilio Ferrara
        and Aram Galstyan %,~\IEEEmembership{Fellow,~IEEE}% <-this % stops a space
\IEEEcompsocitemizethanks{\IEEEcompsocthanksitem Palash Goyal, Homa Hosseinmardi, Emilio Ferrara and Aram Galstyan are with the Department of Computer Science, University of Southern California (USC), and with the USC Information Sciences Institute.
}% <-this % stops an unwanted space
\thanks{Manuscript received April 11, 2018.}}

\markboth{IEEE Transactions on Computational Social Systems,~Vol.~XX, No.~XX, April~2018}%
{Capturing Edge Attributes via Network Embedding}
% \renewcommand{\shortauthors}{B. Trovato et al.}
\IEEEtitleabstractindextext{
\begin{abstract}
Network embedding, which aims to learn low-dimensional representations of nodes, has been used for various graph related tasks including visualization, link prediction and node classification. Most existing embedding methods rely solely on network structure. However, in practice we often have auxiliary information about the nodes and/or their interactions, e.g., content of scientific papers in co-authorship networks, or topics of communication in Twitter mention networks. Here we propose a novel embedding method that uses both network structure and edge attributes to learn better network representations. Our method jointly minimizes the reconstruction error for higher-order node neighborhood, social roles and edge attributes using a deep architecture that can adequately capture highly non-linear interactions. We demonstrate the efficacy of our model over existing state-of-the-art methods on a variety of real-world networks including collaboration networks, and social networks. We also observe that using edge attributes to inform network embedding yields better performance in downstream tasks such as link prediction and node classification.
\end{abstract}

}

% \acmBadgeR{artifacts_available}

\maketitle
\IEEEdisplaynontitleabstractindextext
\input{src/introduction}
\input{src/relatedWork}

\input{src/problem}
\input{src/methodology}

\input{src/exp}

\input{src/results}

\input{src/discussion}
% \newpage
\ifCLASSOPTIONcompsoc
  \section*{Acknowledgments}
\else
  \section*{Acknowledgment}
\fi

This work has been partly funded by the Defense  Advanced  Research  Projects  Agency  (DARPA  \#W911NF-17-C-0094, DARPA \#D16AP00115 and DARPA \#2016-16041100004) and the  Intelligence Advanced Research Projects Activity (IARPA). The views and conclusions contained herein are those of the authors and should not be interpreted as necessarily representing the official policies, either expressed or implied, of DARPA, IARPA, or the U.S. Government. The U.S. Government had no role in study design, data collection and analysis, decision to publish, or preparation of the manuscript. The U.S. Government is authorized to reproduce and distribute reprints for governmental purposes notwithstanding any copyright annotation therein.

\ifCLASSOPTIONcaptionsoff
  \newpage
\fi

\balance
\bibliographystyle{IEEEtran}
\bibliography{bibliography} 

\begin{IEEEbiographynophoto}{Palash Goyal} is a PhD student at the University of Southern California and Research Assistant in the USC Viterbi School of Engineering’s Computer Science Department. His research is funded by IARPA and DARPA. He focuses on analyzing graphs and designing models to understand their behavior.
\end{IEEEbiographynophoto}

\begin{IEEEbiographynophoto}{Homa Hosseinmardi} is a postdoctoral research associate at the Information Sciences Institute, University of South California. She holds a PhD in Computer Science from the University of Colorado Boulder. Her interests lie in the area of computational social science, and mathematical modeling. She is particularly interested in the use of latent factor models and machine learning techniques to study problems with  human behavior, misbehavior and cyber bullying.
\end{IEEEbiographynophoto}

\begin{IEEEbiographynophoto}{Emilio Ferrara} is Assistant Research Professor at the University of Southern California, Research Leader at the USC Information Sciences Institute, and Principal Investigator at the Machine Intelligence and Data Science (MINDS) research group. He is a recipient of the 2016 DARPA Young Faculty Award and of the 2016 Complex System Society Junior Scientific Award. Ferrara's research focuses on machine-learning to model individuals' behaviors and characterize information diffusion in techno-social systems.\end{IEEEbiographynophoto}

\begin{IEEEbiographynophoto}{Aram Galstyan} is a research  director for data science and machine learning at USC's Information Sciences Institute and a research associate professor in the USC Viterbi School of Engineering's Computer Science Department. His current research focuses on different   aspects of machine learning, with applications in computational social sciences, forecasting, and bioinformatics.
\end{IEEEbiographynophoto}
\end{document}

%% file: src/introduction.tex
\IEEEraisesectionheading{\section{Introduction}\label{sec:intro}}
\IEEEPARstart{N}{etworks} exist in various forms in the real world including author collaboration~\cite{Gehrke2003}, social~\cite{freeman2000visualizing}, router communication~\cite{albert2000error}, biological interactions~\cite{theocharidis2009network} and word co-occurrence networks~\cite{i2001small}
Many tasks can be defined on such networks including visualization~\cite{maaten2008visualizing}, link prediction~\cite{liben2007link} and node clustering~\cite{ding2001min} and classification~\cite{bhagat2011node}.
For example, predicting future friendships can be formulated as a link prediction problem on social networks. 
Similarly, predicting page likes by a user and finding community of friends can be regarded as node classification and clustering respectively.
Solving such tasks often involves finding representative features of nodes which are predictive of their characteristics.
Automatic representation of networks in low-dimensional vector space has recently gained much attention and many network embedding approaches have been proposed to solve the aforementioned tasks~\cite{Ahmed2013,Perozzi2014,Cao2015,Tang2015,Grover2016,Ou2016}.

Existing network embedding algorithms either focus on vanilla networks i.e. networks without attributes, or networks with node attributes.
Methods on vanilla networks~\cite{goyal2017graph} obtain the embedding by optimizing an objective function which preserves various properties of the graph.
Of these, many methods~\cite{Ahmed2013,Perozzi2014,Cao2015,Tang2015,Ou2016} preserve proximity of nodes whereas others~\cite{Grover2016,figueiredo2017struc2vec,narayanan2017graph2vec} learn representations which are capable of identifying structural equivalence as well.
%More recent attempts~\cite{huang2017label,huang2017accelerated,chang2015heterogeneous} preserve the proximity of nodes and node labels.
More recent attempts~\cite{huang2017label,huang2017accelerated,chang2015heterogeneous} preserve the proximity of nodes, incorporating both topological structure and node attributes. They learn each embedding separately and align them into a unified space.
However, these approaches fail to incorporate edge attributes present in many real world networks, which can provide more insight into the interactions between nodes. For example, in collaboration networks in which authors are the nodes and edges represent presence of co-authored papers, the content of a co-authored paper can be used to characterize the interaction between the author nodes. 

\begin{figure}
	\centering
	\includegraphics[width=0.99\columnwidth]{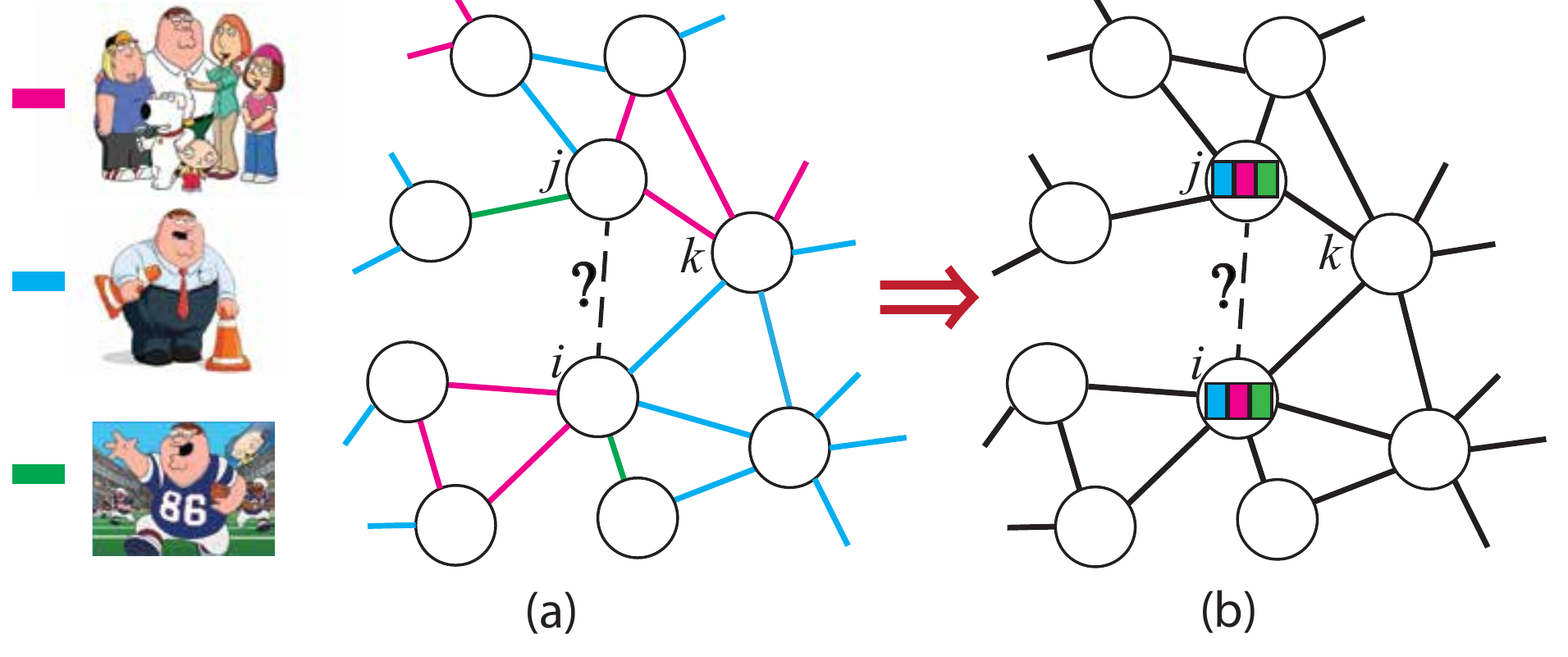}
     \vspace{-.3cm}
	\caption{Users $i$ and $j$ are both engaged in work, family and sport topics. Aggregation of their topics over different interactions, will cause loss of valuable information.}
	\label{fig:aggregation_example}

\end{figure}

Learning representation which can capture node proximity and edge attributes, or labels~\footnote{In this work we use the terms {\em edge attribute} and {\em edge label} interchangeably.}, is a challenging problem.
Although the edge labels can be combined to form node labels, such aggregation incurs loss of information as illustrated by our experiments.
Figure \ref{fig:aggregation_example} shows the effect of this aggregation.
Nodes $i$ and $j$ are both involved in work, family and sport interactions with different people and combining the labels obfuscates the presence of relationship between $i$ and $j$. 
Given that the type of interaction between $i$ and $k$ is ``work" and between $k$ and $j$ is ``family", the likelihood of interaction between $i$ and $k$ is lesser in (a) compared to (b), where $i$ and $j$ have same node labels. 
Another challenge is that the edge labels can be sparse and noisy and the unified representation may fail to capture the heterogeneity of information provided by the the labels and the network.

To overcome the above challenges, in this paper, we introduce \textit{Edge Label Aware Network Embedding} (ELAINE), a model which is capable of utilizing the edge labels to learn a unified representation.
As opposed to linear models, ELAINE uses multiple non-linear layers to learn intricate patterns of interactions in the graph. 
Moreover, the proposed model preserves higher order proximity by simulating multiple random walks from each node and social roles using statistical features and edge labels.
It jointly optimizes the reconstruction loss of node similarity and edge label reconstruction to learn a unified embedding.
% The learned embedding can be used to predict the edge label in addition to the presence of link which is traditionally done in link prediction.
% graph visualization, which visualizes each node in a 2-dimensional space, (b)
We focus our experiments on two tasks: (a) link prediction, which predicts the most likely unobserved edges, and (b) node classification, which predicts labels for each node in the graph.
We compare our model, ELAINE, with the state-of-the-art algorithms for graph embedding.
Furthermore, we show how each component of ELAINE affects the performance on these tasks.
We show results on several real world networks including collaboration networks and social networks. 
Our experiments demonstrate that using a deep model which preserves higher order proximity, social roles and edge labels significantly outperforms the state-of-the-art. % by up to \pgcomment{fill}.
%We also present a case study on French election co-mention network to illustrate the embedding learned by our method.

Overall, our paper makes the following contributions:
\begin{enumerate}
	\item We propose ELAINE, a model for jointly learning the edge label and network structure.
    \item We demonstrate that edge labels can improve performance on link prediction and node classification.
 %   \item We show that ELAINE can be used to predict missing edge labels in networks.
    \item We extend the deep architecture for network representation to preserve higher order proximity and social roles efficiently.
\end{enumerate}

The rest of the paper is organized as follows. 
Section \ref{sec:related} provides a summary of the methods proposed in this domain and differences with our model. 
In Section \ref{sec:problem}, we provide the definitions required to understand the problem and models discussed next. 
We then introduce our model in Section \ref{sec:method}. 
We then describe our experimental setup and obtained results (Sections \ref{sec:exp} and \ref{sec:res}). 
Finally, in Section \ref{sec:discussion} we draw our conclusions and discuss potential applications and future research directions.

%% file: src/relatedWork.tex
\section{Related Work}\label{sec:related}
Generally network embedding techniques come in two flavors: first group uses the pure network structure to map into the embedding space, we call it vanilla network embedding, and the second group combines two sources of information, the topological structure of the graph along with the nodes or link attributes, called attributed network embedding.

\subsection{Vanilla Network Embedding}
There exist variety of embedding techniques for vanilla networks when there is no meta-data available besides the network structure. In general they fall into three broad categories: graph factorization, random-walk based and deep learning based models. Methods such as Locally Linear Embedding~\cite{Roweis2000}, Laplacian Eigenmaps~\cite{belkin2001laplacian}, Graph Factorization~\cite{Ahmed2013}, GraRep~\cite{Cao2015} and HOPE~\cite{Ou2016}, factorize a representative matrix of graph, e.g. node adjacency matrix, to obtain the embedding. Different techniques have been proposed the factorization of the representative matrix based on the matrix properties. Random-walk based techniques are mainly recognized for their power to preserve higher order proximity between nodes by maximizing the probability of occurrence of subsequent nodes in fix length random walk~\cite{Perozzi2014}. node2vec~\cite{Grover2016}, defines a biased random walk to capture structural equivalence, while the structural similarity of nodes is up to the Skip-Gram window size. struc2vec~\cite{figueiredo2017struc2vec} also use a weighted random walk to generate sequence of the structurally similar nodes, independent of their position in network. Recently, deep learning models have been proposed with the ability to capture the non-linear structure in data. SDNE~\cite{Wang2016}, DNGR~\cite{cao2016deep} and VGAE~\cite{kipf2016variational} used deep autoencoders to embed the nodes which capture the nonlinearity in graph.

\subsection{Attributed Network Embedding}
Recently few works have started to use node attributes, beside the network structure in the  embedding process. HNE~\cite{chang2015heterogeneous} embeds multi-modal data of heterogeneous networks into a common space. 
For this purpose, they first apply nonlinear feature transformations on different object types and then with linear transformation project the heterogeneous components into a unified space. 

LANE~\cite{huang2017label} incorporates the node labels into network structure and attributes for learning the embedding representation in a supervised manner. 
They embed attributed network and labels into latent representations separately and then jointly embed them into a unified representation. 
A distributed joint learning process is proposed in~\cite{huang2017accelerated} as an scalable solution, applicable to graphs with large number of nodes and edges. 

%More recent attempts~\cite{huang2017label,huang2017accelerated,chang2015heterogeneous} preserve the proximity of nodes and node labels. They learn each embedding separately and align them into a unified space.  However this aggregation will cause the loss of information about the type of interaction among a node with each of its neighbors. 
Edge attributes can be aggregated and assigned to nodes and provided as input into these methods. 
However, this aggregation will cause the loss of information about the type of interaction between a node and its neighbors. 
To overcome this challenge we introduce a model to utilize the edge labels in the process of mapping nodes in a unified space. 

Note finally that an alternative approach to network embedding is provided via generative graphical models such as the Mixed Membership Stochastic Blockmodel~\cite{Airoldi2008}, and its extension that take into account node attributes~\cite{Nallapati2008,Chang2009,Cho2016}.

%% file: src/problem.tex
\section{Problem Statement}\label{sec:problem}

We denote a weighted graph as $G(V, E)$ where $V$ is the vertex set and $E$ is the edge set. 
The weighted adjacency matrix of $G$ is denoted by $A$. 
If $(i,j)\in E$, we have $A_{ij} > 0$ denoting the weight of edge $(i,j)$; otherwise we have $A_{ij}=0$. 
We use $\vct{a_i} = [A_{i,1}, \cdots, A_{i, n}]$ to denote the $i$-th row of the adjacency matrix. 
We use $E^a \in \mathbb{R}^{m \times p}$ to denote the edge attribute matrix and $\vct{e^a_{ij}} = [e^a_{ij1}, \cdots, e^a_{ijp}]$ to denote the attributes of edge $(i, j)$, where $p$ is the number of edge attributes.

We define our problem as follows: \textit{Given a graph $G=(V,E)$ and associated edge attributes $E^a$, we aim to represent each node $u$ in a low-dimensional vector space $\vct{y_u}$ by learning a mapping $f:\{V, E^a\} \rightarrow \mathbb{R}^d$, namely $\vct{y_v}=f(v, E^a) \ \forall v \in V$}. 
We require that $d \ll n$ and the function $f$ preserves some proximity measure defined on the graph $G$. 
Intuitively, if two nodes $u$ and $v$ are ``similar'' in graph $G$, their embedding $\vct{y_u}$ and $\vct{y_v}$ should be close to each other in the embedding space. 
We use the notation $f(G)\in \mathbb{R}^{n\times d}$ for the embedding matrix of all nodes in the graph $G$. 
Note that the embedding of an edge $(u, v)$ is defined as $g(u, v) = [\vct{y_u}, \vct{y_v}]$, i.e. the concatenation of embeddings of nodes $u$ and $v$. 
It can be written as $g:E \rightarrow \mathbb{R}^{2d}$. 
We use $g(u, v)$ to reconstruct the edge label $\vct{e^a_{uv}}$. This enables us to infer the missing edge labels by using the adjacency of the incident nodes.

%% file: src/methodology.tex
\section{ELAINE}\label{sec:method}
We propose an edge label aware information network embedding method - ELAINE, which models $l^{th}$-order proximity, social role features and edge labels using a deep variational autoencoder.
The core component of the model is based on a deep autoencoder which can be used to learn the network embedding by minimizing the following loss function:

\begin{equation}
L = \sum_{i=1}^n \|(\vct{\hat{a}_i} - \vct{a_i}) \odot \vct{\beta_i}\|_2^2 = \|(\hat{A} - A) \odot \mathcal{B}\|_F^2
\end{equation}
\begin{figure}
    \centering
    \includegraphics[width=0.22\textwidth]{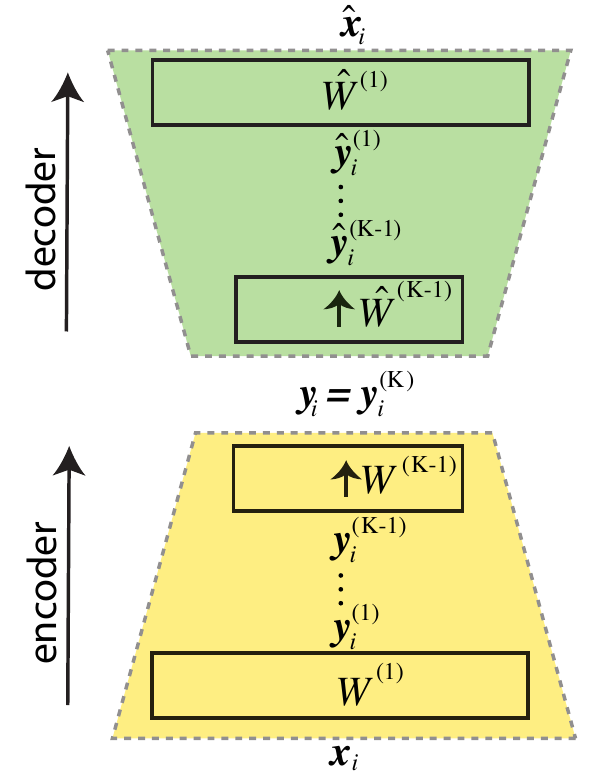}
    \caption{Traditional deep autoencoder model.}
    \label{fig:sdne}
    \vspace{-.5cm}
\end{figure}

%AutoEmbed
Figure \ref{fig:sdne} illustrates the autoencoder.
The objective function penalizes inaccurate reconstruction of node neighborhood.
As many legitimate links are not observed in the networks, a weight $\vct{\beta_i}$ is traditionally used to impose more penalty on reconstruction of observed edges~\cite{Wang2016}.

Although the above model can learn network representations which can reconstruct the graph well, it suffers from four challenges.
Firstly, as the model reconstructs the observed neighborhood of each vertex, it only preserves second order proximity of nodes. 
\emph{Wang et. al.}~\cite{Wang2016} extend the model to preserve first order proximity but their model fails to capture higher order proximities. 
Concretely, if two nodes have disjoint neighborhoods the model will keep them apart regardless of the similarity of their neighborhoods.
Secondly, the model is prone to overfitting leading to a satisfactory reconstruction performance but sub-par performance in tasks like link prediction and node classification.
\emph{Wang et. al.}~\cite{Wang2016} use $l_1$ and $l_2$ regularizers to address this issue but we show that using variational autoencoders can achieve better performance.
Thirdly, the model does not explicitly capture social role information.
Real world networks often have a role based structure understanding which can help with various prediction tasks.
Lastly, the model does not consider edge labels. 
We show that incorporating edge label reconstruction leads to improved performance in various tasks.

\begin{figure}
	\centering
	\includegraphics[width=0.51\textwidth]{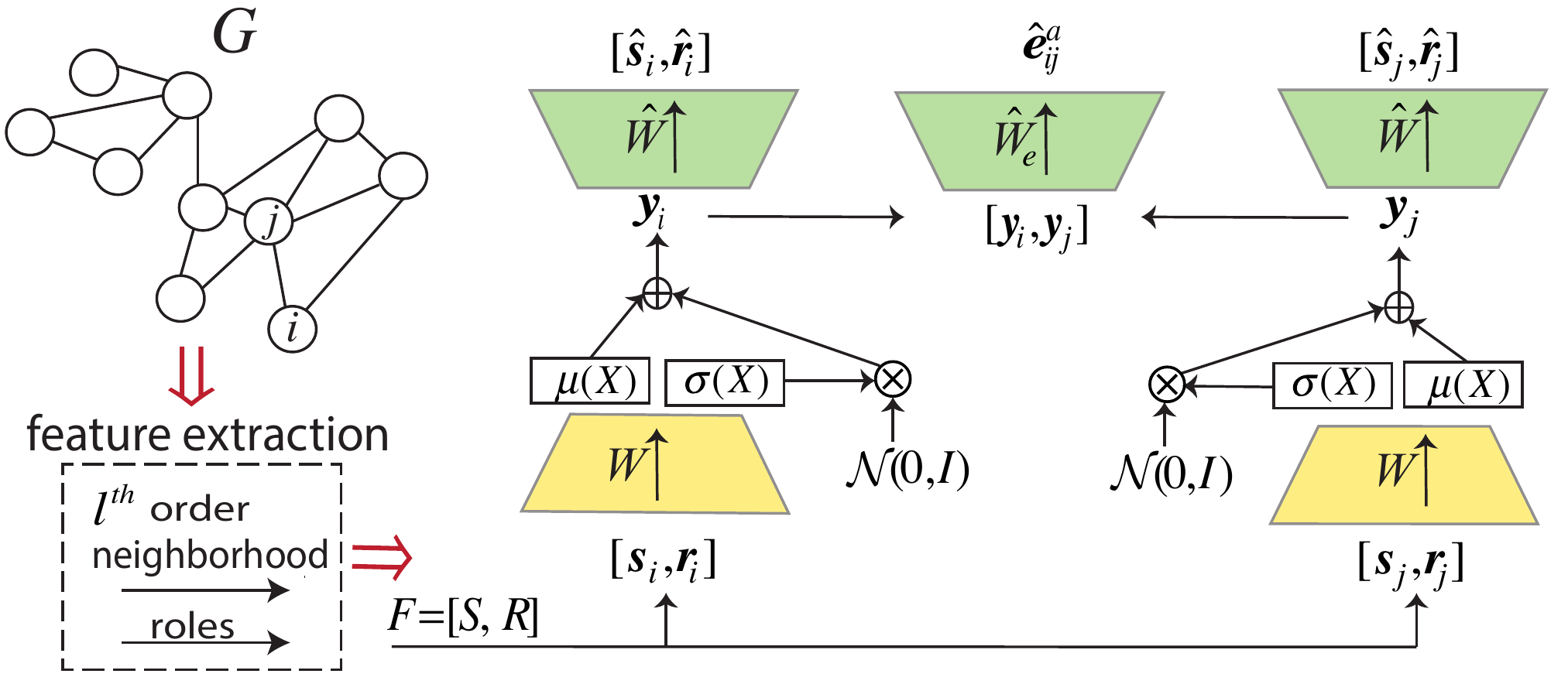}
     \vspace{-.3cm}
	\caption{Edge label aware embedding model. ELAINE extracts higher order relations between nodes using random walks and social role based features. The coupled autoencoder jointly optimizes these features and edge attributes to obtain a unified representation.}
	\label{fig:model}
	\vspace{-.5cm}
\end{figure}

To address the above challenges, we propose a random walk based deep variational autoencoder model with an objective to jointly optimize the higher order neighborhood, role based features and edge label reconstruction.

\begin{figure*}
	\centering
	\includegraphics[width=0.85\textwidth]{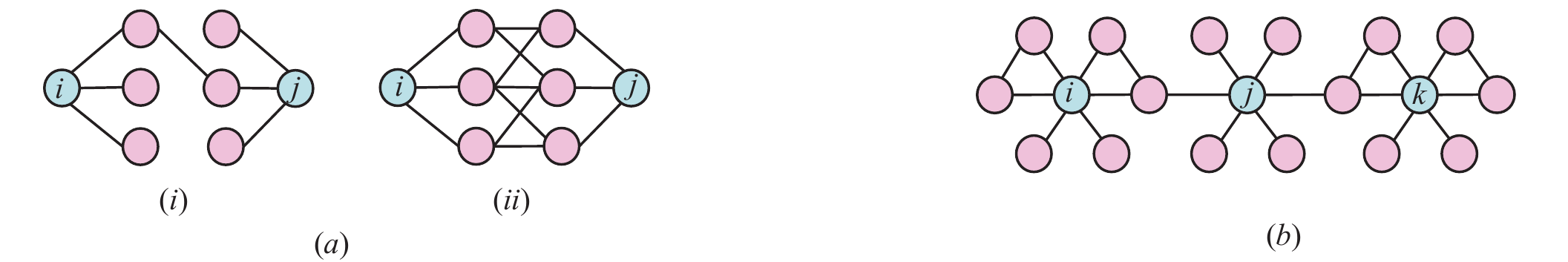}
     \vspace{-.3cm}
	\caption{Importance of capturing higher order proximity and social roles. (a) Nodes $i$ and $j$ are more similar in (ii) compared to (i) but first order proximity fails to capture this, (b) Node $i$ and $k$ have similar roles, but they are far apart in network. Using social role indicative statistical features can capture similarity of these nodes.}
	\label{fig:features}
	\vspace{-.2cm}
\end{figure*}

\subsection{Variational Autoencoder}
As we aim to find a low-dimensional manifold the original graph lies in, we want to learn a representation which is maximally informative of observed edges and edge labels.
At the same time, as the autoencoder penalizes reconstruction error, it encourages perfect reconstruction at the cost of overfitting to the training data.
This is in particular problematic for learning representations for graphs as networks are constructed from interactions which may be incomplete or noisy.
We want to find embeddings which are robust to such noise and can help us in tasks such as link prediction and node classification.

Many methods have been proposed to improve the generalization of autoencoders for tasks like image and speech recognition \cite{bengio2013representation}.
Of these, sparse autoencoders \cite{boureau2008sparse}, which use $L_1$ and $L_2$ penalty on weights, and stacked denoising autoencoders \cite{vincent2010stacked}, which sample autoencoder inputs by adding Gaussian noise to data inputs, have been shown to improve performance in graph related tasks \cite{Wang2016,cao2016deep}. 
Nonetheless, these models suffer from various challenges. The former doesn't ensure a smooth manifold and the latter is sensitive to the number of corrupted inputs generated \cite{rifai2011contractive}.
We propose to use variational autoencoder for graphs and illustrate in Section \ref{sec:res} that it can improve performance in different tasks.

Variational autoencoders (VAEs) look at autoencoders from a generative network perspective. 
The model aims to maximize $P(X) = \int P(X|z; \theta)P(z)dz$, where $X$ is the training data, $z$ is the latent variable.
They assume $P(X|z; \theta)$ to be normally distributed, i.e. $P(X|z; \theta) = \mathcal{N}(X|f(z;\theta), \sigma^2*I)$, where $f(z;\theta)$ is the decoding of $z$ with the learned decoder parameters $\theta$.
Computing the integral is intractable and is approximated by summation.
Moreover, for most $z$, $P(X|z)$ will be nearly zero and thus we need to find $z$ which are more likely, given the data.
This can be written as finding the distribution $Q(z|X)$ which is approximated using the encoder enabling us to compute $E_{z\sim Q}P(X|z)$ tractably.
The model assumes a normal form for $Q(z|X)$.
Thus, we have $Q(z|X) = \mathcal{N}(z|\mu(X; \phi), \Sigma(X; \phi))$, where $\phi$ are parameters of the encoder.
Typically, $\Sigma$ is constrained to be a diagonal matrix to decouple the latent variables.
The optimization is reduced to $E_{z\sim Q}[log P(X|z)] -D_{KL}[Q(z|X)|| P(z)]$.
The second term is the KL-divergence between two multivariate Gaussian distribution and the first term is the likelihood of reconstruction given the latent variables constrained on their distributions learned by the autoencoder.

In practice, VAEs can be trained by minimizing the sum of two terms: (1) reconstruction loss and (2) KL-divergence of latent variable distribution and unit Gaussian, using backpropagation.
The variance of reconstruction controls the generalization of the model which can be treated as the coefficient of KL-divergence loss.

\subsection{Higher Order Proximity and Role Preservations}

Nodes in a network are related to each other via many degrees of connection.
Some nodes have direct connections while others are connected through paths of varying lengths.
Moreover, nodes may take several different roles.
For e.g., in web graphs, nodes can be broadly classified to hubs and authorities.
Hubs refer to nodes which refer to other nodes i.e. have high out-degree whereas authorities refer to nodes which are linked to by other nodes.
A good embedding should preserve such higher order and role based relations between nodes.
Naively using node adjacency as the input, an autoencoder cannot achieve this as shown in Figure \ref{fig:features} (a).
Nodes $i$ and $j$ have different neighborhoods and the model cannot differentiate between (i) and (ii).
In both cases, the model will keep them far apart although in (ii), the nodes are more similar.
Similarly, in Figure \ref{fig:features} (b), we see that nodes $i$ and $k$ are structurally similar but proximity based methods cannot utilize this.

\subsubsection{Random Walks}
To preserve higher order proximities, we obtain global distance based similarities of each node with the rest of the nodes.
One way to obtain such a set of vectors is to use metrics such as Katz Index \cite{katz1953new}, Adamic Adar \cite{adamic2003friends} and Common Neighbors \cite{newman2001clustering}.
Although such metrics capture global proximities accurately, their computation is inefficient and the time complexity is up to $O(n^3)$.
We overcome the inefficiency by approximating them using random walks \cite{backstrom2011supervised}.
For each node $i$, we simulate $k$ random walks each of length $l$.
Each random walk, $\lbrace v_{i,1}, v_{i,2} \ldots v_{i,l}\rbrace$, from node $i$ generates a node $j$ with probability:
\[
    P(v_{i,j}|v_{i,j-1})=\left\{
                \begin{array}{ll}
                  \frac{1}{d_{j-1}}\ \textnormal{if} \ (v_{i,j-1}, v_{i,j}) \in E\\
                  0\ otherwise\\
                \end{array}
              \right.
  \]
where $d_k$ is the degree of node $k$.
Note that since a random walk of length $l$ from node $i$ is equivalent to a random walk of length $l-1$ for node $v_{i, 1}$, generating $k$ random walks of length $l$ only requires $O(k)$ time each node.

\subsubsection{Role preserving features}
Social roles in a network are characterized by various local and global statistics. 
For example, high degree can be reflective of social importance.
Broadly, we classify role discriminating features into two categories: (a) statistical features, and (b) edge attributes.
We consider the following statistical features which have been shown to correlate with social roles\cite{henderson2012rolx}: (i) node's degree, (ii) weighted degree, (iii) clustering coefficient, (iv) eccentricity, (v) structural hole and (vi) local gatekeeper. 
We append these features with node's neighborhood as input to our model.
Having such statistical features helps obtain an embedding which preserves social roles.
On the other hand, a node can take different roles with different neighbors (henceforth referred as interactive roles) which cannot be captured by such statistical features. For example, in a collaboration network, author $i$ may take the role of Professor with his student $j$ and colleague with another professor $k$. Identifying such distribution of roles can help model the network more accurately. For this we use the edge attributes which can be reflective of such interactions. Concretely, we consider the topics of conversation between nodes and jointly optimize their reconstruction of node neighborhood reconstruction.

\subsection{Incorporating edge labels}
Autoencoder defined above takes node neighborhood and statistical role preserving features as input and aims to reconstruct them.
One possible approach to incorporate edge attributes is to aggregate them for each node  and append them with other node features.
The drawback of this approach is that information loss can incur following aggregation.
Such aggregation cannot preserve interactive roles between nodes.

We propose to overcome this problem by coupling copies of autoencoders for nodes $i$ and $j$. The model is composed of a coupled autoencoder and an edge attribute decoder, Figure \ref{fig:model}.
The intuition is to force the embeddings of nodes $i$ and $j$ to capture information pertaining to the attributes of the edge between them.
This is ensured by adding the edge attribute reconstruction loss to the objective function.
Thus, we learn model parameters by minimizing a loss function with the following terms:
\subsubsection{Neighborhood and social role reconstruction} The $l$th-order neighborhood of each node along with the social role preserving statistical features:
\begin{align*}
	L_{n} &= \|([\hat{S}, \hat{R}] - [S, R]) \odot \mathcal{B}\|_F^2,
\end{align*}
where each row of $S \in \mathbb{R}^{n\times n}$ and $R \in \mathbb{R}^{n\times r}$ compose of neighborhood similarity and role statistics respectively. Henceforth, we will refer to $[\hat{S}, \hat{R}]$ by $\hat{F} \in \mathbb{R}^{n\times n+r}$ and $[S, R]$ by $F \in \mathbb{R}^{n\times n+r}$.

\subsubsection{Edge label/attributes reconstruction}
For each pair of nodes, we reconstruct the attributes of the edge between them:
\begin{align*}
	L_{e} &= \|\hat{E^a} - E^a\|_F^2,
\end{align*}
where each row $i$ of $E^a \in \mathbb{R}^{m\times p}$ is the vector of attributes of the $i$th edge .

\subsubsection{Regularization}
To avoid overfitting, we use three types of regularizations: (a) Lasso ($L_l$), (b) Ridge ($L_r$), and (c) Variational loss ($L_v$), defined below:
\begin{align*}
	L_{v} &= D_{KL}(Q(z|X)||P(z)),\\
    L_l &= \sum_{k=1}^K \left( \|W^{(k)}\|_{sum} + \|\hat{W}^{(k)}\|_{sum}  + \|W_{e}^{(k)}\|_{sum}\right),\\
	L_r &= \sum_{k=1}^K \left( \|W^{(k)}\|_F^2 + \|\hat{W}^{(k)}\|_F^2 + \|\hat{W_e}^{(k)}\|_F^2\right),\\
    L_{reg} &= \alpha_v L_{v} + \alpha_l L_l + \alpha_r L_r,
\end{align*}
where $Q(z|X)$ corresponds to the encoder and $P(z)$ is the prior which is assumed to be unit Gaussian.
\begin{algorithm}[tb]
\Fn{ELAINE (Graph $G=(V,E)$, Edge attributes $E^a \in \mathbb{R}^{m\times p}$, Dimensions $d$, Random walk parameters $rw_{param}$)}{
 $S$ $\leftarrow$ RandomWalk($G$, $rw_{param}$)\;
 $R$ $\leftarrow$ GetSocialRoles($G$)\;
 $F$ $\leftarrow$ [$S$, $R$]\;
 $\vartheta$ $\leftarrow$ RandomInit()\;
 Set $\mathcal{F} = \{(f_i, f_j, e^a_{ij})\}$ for each $e=(v_i,v_j) \in E$, $f_i, f_j \in F$\;
 \For{$iter=1 \ldots MAX\_ITER$}{
 	Randomly sample minibatch $M$ from $\mathcal{F}$\;
 	$L = L_{n} + \alpha_1 L_{e} + L_{reg}$\;
    $grad$ $\leftarrow$ $\partial L/ \partial \vartheta$\;
    $\vartheta$ $\leftarrow$ UpdateGradAdam($\vartheta$, $grad$)\;
 }
 $Y$ $\leftarrow$ EncoderForwardPass($G$, $\vartheta$)\;
 return $Y$}
\caption{ELAINE}
\label{alg:elaine}
\end{algorithm}
The overall objective function thus becomes the following:
\begin{align}\label{eqn:loss_fn}
	L &= L_{n} + \alpha_1 L_{e} + L_{reg},
\end{align}

\begin{table}[!htbp]
	\centering\footnotesize
    \renewcommand{\arraystretch}{1.3}
    \caption{Dataset Statistics}
    \label{tab:data_summ}
    \vspace{-10pt}
  \begin{tabular}{ l | c | c | c | c  }
  %\begin{tabular}{| p{1.9cm} | l | p{1.5cm} | p{1.5cm} |}
           \hline \hline
      Name & Hep-th & Twitter & 20-new group & Enron\\ \hline \hline
      $n$ & 7,980 & 6,479 & 1,727 & 145\\ \hline
      $m$ & 21,036 & 18,123 & 2,980,802 & 912\\ \hline
      Avg. degree & 5.27 & 5.59 & 1726 & 12.58\\ \hline
%       Diameter & 3 & 4.59 & 64.78 & 8.54 & 5.27 & 31.55 & 1,719 & 19.91\\ \hline
      \# of node labels & 20 & - & 3 & -\\ \hline
      \# of edge attributes & 100 & 10 & 6 & 10\\ \hline \hline
  \end{tabular}
\end{table}

\subsection{Optimization}
% \begin{algorithm}
% \Fn{ELAINE (Graph $G=(V,E)$, Edge attributes $E^a \in \mathbb{R}^{m\times p}$, Dimensions $d$, Random walk parameters $rw_{param}$)}{
%  $S$ $\leftarrow$ RandomWalk($G$, $rw_{param}$)\;
%  $R$ $\leftarrow$ GetSocialRoles($G$)\;
%  $F$ $\leftarrow$ [$S$, $R$]\;
%  $\theta$ $\leftarrow$ RandomInit()\;
%  Set $\mathcal{F} = \{(f_i, f_j, e^a_{ij})\}$ for each $e=(v_i,v_j) \in E$
%  \While{$iter$ $<$ MAX_ITER}{
%  	Randomly sample minibatch $M$ from $\mathcal{F}$\;
%     L &= L_{n} + \alpha_1 L_{e} + L_{reg}\;
%     $grad$ $\leftarrow$ $\partial L/ \partial \theta$
%     $\theta$ $\leftarrow$ UpdateGradAdam($\theta$, $grad$)\;
%  	$iter$++\;
%  }
%  $Y$ $\leftarrow$ EncoderForwardPass($G$)\;
%  return $Y$}
% \caption{ELAINE}
% \end{algorithm}

To get the optimal parameters for the model defined above, we minimize the loss function $L$.
The optimization involves three sets of gradients: $\partial L/\partial W^{(k)}$, $\partial L/\partial \hat{W}^{(k)}$ and $\partial L/\partial \hat{W_e}^{(k)}$.
Applying the gradients on equation \ref{eqn:loss_fn}, we get:
\begin{align*}
	\frac{\partial L}{\partial W^{(k)}} &= \frac{\partial L_n}{\partial W^{(k)}} + \alpha_1 \frac{\partial L_e}{\partial W^{(k)}} + \frac{\partial L_{reg}}{\partial W^{(k)}},\\
    \frac{\partial L}{\partial \hat{W}^{(k)}} &= \frac{\partial L_n}{\partial \hat{W}^{(k)}} + \frac{\partial L_{reg}}{\partial \hat{W}^{(k)}},\\
    \frac{\partial L}{\partial \hat{W_e}^{(k)}} &= \alpha_1 \frac{\partial L_e}{\partial \hat{W_e}^{(k)}} + \frac{\partial L_{reg}}{\partial \hat{W_e}^{(k)}},
\end{align*}
where $k = \lbrace 1,2, \ldots, K \rbrace$. For $k=K$, we have 
\begin{align*}
	\frac{\partial L_n}{\partial \hat{W}^{(K)}} &= [2 (\hat{F} - F) \odot \mathcal{B}][ \frac{\partial \ddot{a}(\hat{Y}^{(K-1)}\hat{W}^{(K)} + \hat{b}^{(K)})}{\partial \hat{W}^{(K)}}],\\
    \frac{\partial L_e}{\partial \hat{W_e}^{(K)}} &= [2 (\hat{E^a} - E^a)][ \frac{\partial \ddot{a}(\hat{E^a}^{(K-1)}\hat{W_e}^{(K)} + \hat{b}^{(K)})}{\partial \hat{W_e}^{(K)}}],
\end{align*}
where $\ddot{a}()$ represents the activation function of the autoencoder. We use the above derivatives and backpropagate them to get the derivatives for other $k$ values and $\partial L_{.}/ \partial W^{(k)}$ for each $L_{.}$.

After obtaining the derivatives we optimize the model using stochastic gradient descent (SGD) \cite{rumelhart1988neurocomputing} with Adaptive Moment Estimation (Adam)\cite{kingma2014adam}.
The algorithm is specified in Algorithm \ref{alg:elaine}.

%% file: src/exp.tex
\section{Experiments}\label{sec:exp}
In this section, we first describe the data sets used and then discuss the baselines we use to compare our model.

This is followed by the evaluation metrics for our experiments and parameter settings.
All the experiments were performed on a Ubuntu 14.04.4 LTS system with 32 cores, 128 GB RAM and clock speed of 2.6 GHz.
The GPU used was Nvidia Tesla K40C.

\begin{figure*}[!htbp]
	\centering
	\includegraphics[width=0.99\textwidth]{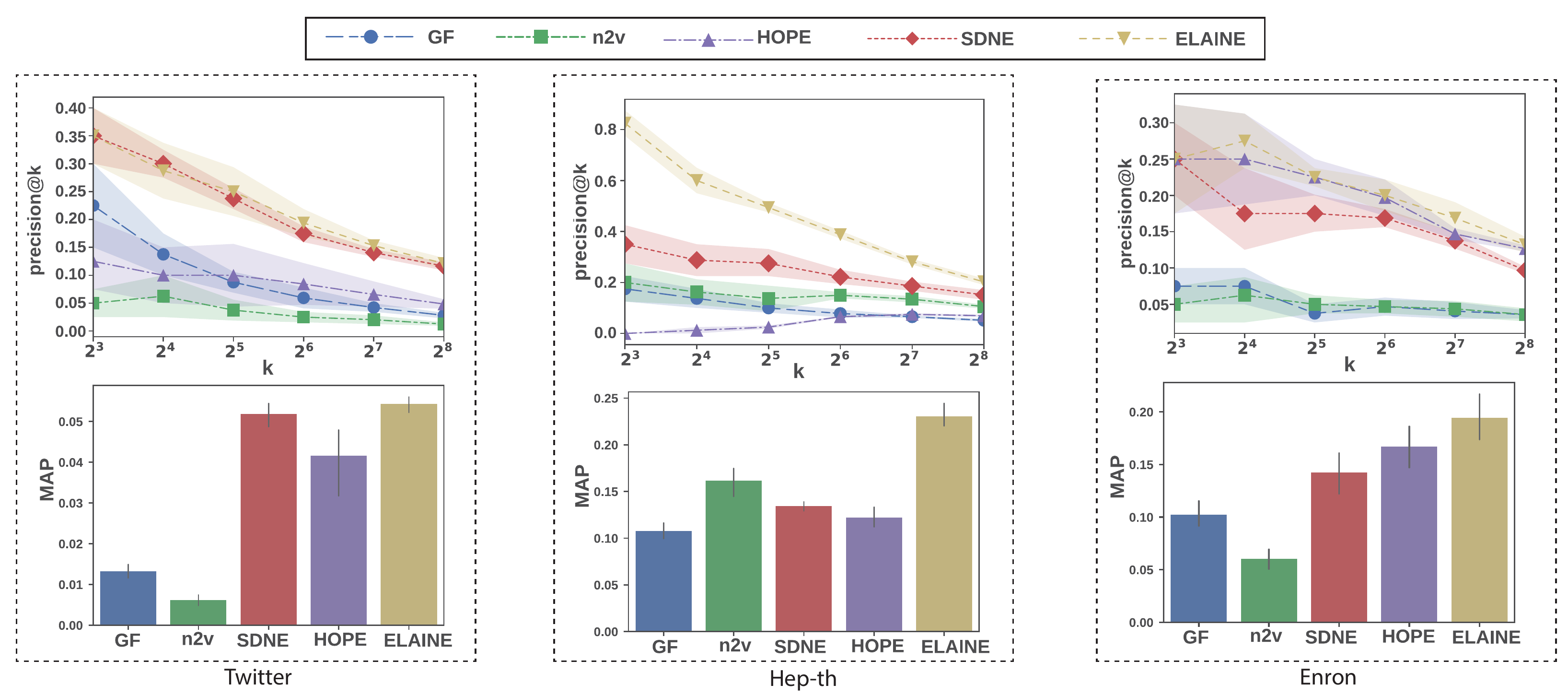}
     \vspace{-.3cm}
	\caption{Precision@k and MAP of link prediction for different data sets.}
	\label{fig:LP}
    
\end{figure*}

\subsection{Datasets}

We conduct experiments on four real-world datasets to evaluate our proposed algorithm.
The datasets are summarized in Table \ref{tab:data_summ}.

\textbf{Hep-th}~\cite{Gehrke2003}: The original data set contains abstracts of papers in High Energy Physics Theory conference in the period from January 1993 to April 2003. 
We create a collaboration network for the first five years.
We get the node labels using the Google Scholar API \footnote{\url{https://pypi.python.org/pypi/scholarly/0.2}} to obtain university labels for each author.
We apply NMF \cite{sra2006generalized} on the set of abstracts to get topic distribution for each abstract.
We aggregate the topic distribution of all the coauthored papers between two authors to get the edge attributes.

\textbf{Twitter}~\cite{ferrara2017disinformation}: The data set consists of tweets on the French election day, 7th May, 2017.
The tweets were obtained using keywords related to election including \textit{France2017}, \textit{LePen}, \textit{Macron} and \textit{FrenchElections}.
We construct the mention network by connecting users who mention each other in a tweet.
The topic distribution of the tweet between user $i$ and $j$, obtained from NMF,  is regarded as the edge attribute $\textit{\textbf{e}}^a_{ij}$.
%For node labels, we use the tag which indicates if the user's profile will be deactivated after the election.

\textbf{20-Newsgroup}\footnote{\url{http://qwone.com/~jason/20Newsgroups/}}: This dataset contains about 20,000 newsgroup documents each corresponding to one of 20 topics. 
For our experiments, we selected all documents in three news group  ``computer graphics", ``sport-baseball" and ``politics and guns". 
From this we construct a document similarity graph using the cosine similarity of their tf-idf vectors.
Similar to Hep-th, we use NMF to get topic distribution of each document and use common topics between documents as the edge attributes.

%\footnote{\url{https://www.cs.cmu.edu/~./enron/}}
\textbf{Enron~\cite{Enron2004}}: This dataset contains emails communicated among about 150 users, mostly senior management of Enron. We connect two users if they have exchanged an email. Edge attribute between node $i$ and $j$ is the extracted topics from each set of emails between them using NMF.

\subsection{Baselines}
We compare our model with the following state-of-the-art methods:
\begin{itemize}
% 	\item \textit{Locally Linear Embedding} \cite{Roweis2000}: It assumes every node is a linear combination of its neighbors and minimizes the reconstruction loss of getting a node's embedding from its neighbors'.
% 	\item \textit{Laplacian Eigenmap}s \cite{belkin2001laplacian}: It penalizes the weighted square of distance between neighbors. This is equivalent to factorizing the normalized Laplacian matrix.  
	\item \textit{Graph Factorization} (GF) \cite{Ahmed2013}: It factorizes the adjacency matrix with regularization.
    \item \textit{Structural Deep Network Embedding} (SDNE) \cite{Wang2016}: It uses deep autoencoder along with Laplacian Eigenmaps objective to preserve first and second order proximities.
    \item \textit{Higher Order Proximity Preserving} \cite{Ou2016} (HOPE): It factorizes the higher order similarity matrix between nodes using generalized singular value decomposition \cite{paige1981towards}.
    \item \emph{node2vec} \cite{Grover2016}: It preserves higher order proximity by maximizing the probability of occurrence of subsequent nodes in fixed length biased random walks. They use shallow neural networks to obtain the embeddings. \emph{DeepWalk} is a special case of \emph{node2vec} with the random walk bias set to 0.
\end{itemize}

\subsection{Evaluation Metrics}\label{subsec:metrics}
In our experiments, we evaluate our model on tasks of link prediction, node classification and visualization.
For link prediction, we use precision@$k$ and Mean Average Precision (MAP) as our metric.
For node classification, we use $micro-F1$ and $macro-F1$.

\begin{figure*}[!htbp]
\includegraphics[width=1.0\textwidth]{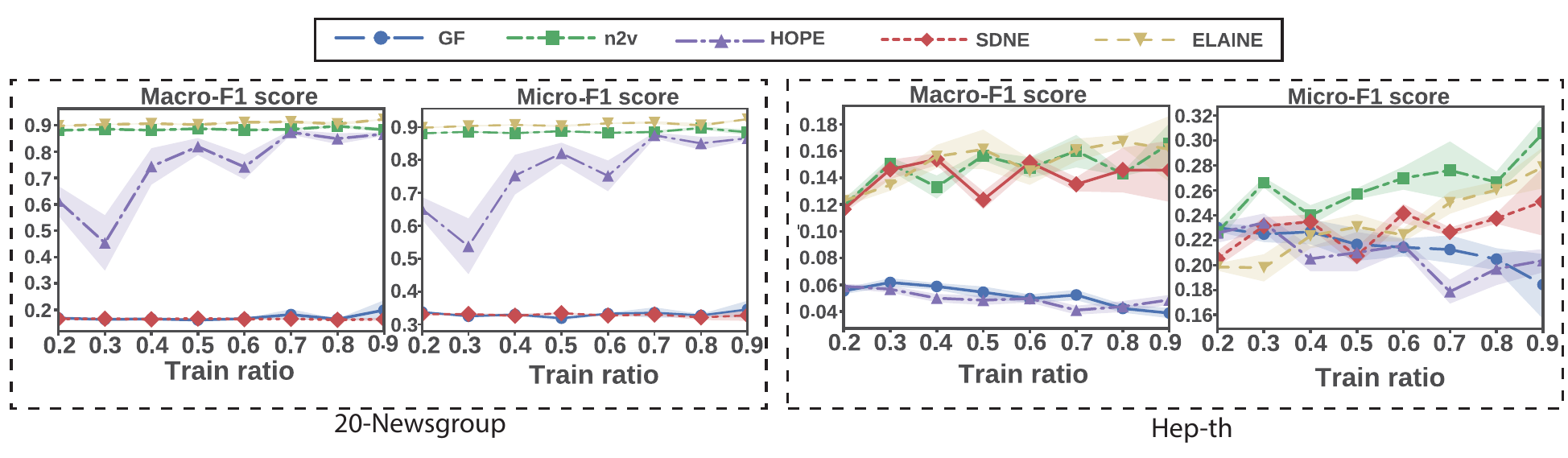}
 \vspace{-0.7cm}
\caption{Node classification results for different data sets.}
\label{fig:nc}

\end{figure*}

The formulae used for these metrics are as follows:

\textbf{precision@$k$}: It is the fraction of correct predictions in top $k$ predictions. It is defined as $\frac{|E_{pred}(k) \cap E_{gt}|}{k}$, where $E_{pred}$ and $E_{gt}$ are the predicted and ground truth edges respectively.
        
\textbf{MAP}: It averages the precision over all nodes. It can be written as $MAP = \frac{\sum_i AP(i)}{|V|}$, where $AP(i) = \frac{\sum_k precision@k(i) \cdot \mathbb{I}\{E_{pred_i}(k) \in E_{gt_i}\}}{|\{k: E_{pred_i}(k) \in E_{gt_i}\}|}$ and $precision@k(i) = \frac{|E_{pred_i}(1:k) \cap E_{gt_i}|}{k}$
        
\textbf{\textit{macro-F1}}, in a multi-label classification task, is defined as the average $F1$ of all the labels, i.e., $macro-F1 = \frac{\sum_{l \in \mathcal{L}} F1(l)}{|\mathcal{L}|}$, where $F1(l)$ is the $F1$-score for label $l$.

\textbf{\textit{micro-F1}} calculates $F1$ globally by counting the total true positives, false negatives and false positives, giving equal weight to each instance. It is thus $\frac{2*P*R}{P+R}$, where $P$ and $R$ are overall precision and recall respectively.

\subsection{Parameter settings}
In our experiments, we use two hidden layers for feature encoder and decoder with size [500, 300].
For the edge attribute decoder, we experiment with a single hidden layer with 1000 neurons and without any hidden layer.
% In general, we find that not using a hidden layer achieves best performance.
Optimal values of other hyperparameters such as $\alpha_1$, $\alpha_r$, $\alpha_l$ and $\alpha_v$ are obtained using grid search over $[10^{-5}, 10^3]$ in factors of 10.

%% file: src/results.tex
\section{Results and Analysis}\label{sec:res}
In this section, we present results of our model on link prediction and node classification, and provide a comparison with baselines.
Moreover, we discuss the effect of each component of our model to the overall precision gain.
We then discuss the sensitivity of our model to different hyperparameters.

\subsection{Link Prediction}

\iffalse
\begin{figure*}[!htbp]
	\centering
	\subfloat[Hep-th]{\label{fig_m2} \includegraphics[width=0.32\textwidth]{img/HepthEdgeLabel_lp_d_128}}
	\hfil
	\subfloat[Twitter]{\label{fig_m2} \includegraphics[width=0.32\textwidth]{img/TwitterFrenchEl_lp_d_128}}
	\hfil
	\subfloat[Enron]{\label{fig_m2} \includegraphics[width=0.32\textwidth]{img/Enron_lp_d_20}}
	\hfil
	\caption{Precision@k of link prediction for different data sets.}
	\label{fig:p_at_k_128_lp}
    \vspace{-.5cm}
\end{figure*}
\begin{figure*}[!htbp]
	\centering
	\subfloat[Hep-th]{\label{fig_m2} \includegraphics[width=0.32\textwidth]{img/HepthEdgeLabel_lp_d_128_map}}
	\hfil
	\subfloat[Twitter]{\label{fig_m3} \includegraphics[width=0.32\textwidth]{img/TwitterFrenchEl_lp_d_128_map}}
	\hfil
	\subfloat[Enron]{\label{fig_m4} \includegraphics[width=0.32\textwidth]{img/Enron_lp_d_20_map}}
	\hfil
	\caption{MAP of link prediction for different data sets.}
	\label{fig:map_128_lp}
    \vspace{-.5cm}
\end{figure*}
\fi

Information networks are meant to capture the interactions in real world.
This translation of interactions can be noisy and inaccurate.
Predicting missing links in the constructed networks and links likely to occur in the future is an important and difficult task.
We test our model on this link prediction task to understand the generalizability of our model.
For each network, we randomly hide 20\% of the network edges.
We use the rest of the network to learn the embeddings of nodes and sort the likelihood of each unobserved edge to predict the missing links.
As number of node pairs for a network of size $N$ is $N(N-1)/2$, we randomly sample 1024 nodes for evaluation (similar to \cite{goyal2017graph}).
We get 5 samples for each data set and report the mean and standard deviation of precision and MAP values.

% \ref{fig:p_at_k_128_lp} and \ref{fig:map_128_lp}
Figures \ref{fig:LP} illustrates the link prediction $precision@k$ and MAP values for the methods on data sets.
We observe that our model significantly outperforms baselines on Hep-th.
This implies that using the topic distribution of abstracts can help us understand the relation between authors.
On Twitter and Enron, we observe that gain in $precision@k$ isn't as significant as gain in MAP.
Thus, our model improves predictions considerably for nodes with lesser incident edges although the top predicted edges are slightly better than baselines.
This follows intuition since edge labels for such nodes provide higher information about their relation with other nodes than for the nodes for which we have ample edge information.
We also observe that our model achieves higher improvement over baselines on Hep-th and Enron compared to Twitter.
This can be attributed to the characteristic of tweets which tend to be more unstructured and noisy and hence more challenging to model.
Overall, gain in performance consistently for different kinds of data sets shows that our model can utilize edge attributes in different domains and improve link prediction performance.

\subsection{Node Classification}
\begin{table*}[!htbp]
	\centering\footnotesize
    \renewcommand{\arraystretch}{1.3}
    \caption{Common interests of authors in Hep-th}
    \label{tab:hepth_labels}
    \vspace{-10pt}
%   \begin{tabular}{@{}c@{}|@{}c@{}|@{}c@{}|@{}c@{}|@{}c@{}|@{}c@{}|@{}c@{}}
\begin{tabular}{|c|c|c|c|c|}
  %\begin{tabular}{| p{1.9cm} | l | p{1.5cm} | p{1.5cm} |}
           \hline \hline
      \textbf{string theory} & \textbf{theoretical physics} & \textbf{physics} & \textbf{quantum field theory} & \textbf{mathematical physics}\\ \hline
      21.64\% & 19.65\% & 16.17\% & 15.17\% & 13.68\% \\ \hline
      \textbf{cosmology} & \textbf{quantum gravity} & \textbf{particle physics} & \textbf{high energy physics} & \textbf{machine learning}\\ \hline
      9.20\% & 7.21\% & 7.21\% & 6.22\% & 4.48\%\\ \hline
      \textbf{supersymmetry} & \textbf{black holes} & \textbf{bioinformatics} & \textbf{gravity} & \textbf{noncomm. geometry}\\ \hline
      4.23\% & 4.23\% & 4.23\% & 3.98\% & 3.98\%\\ \hline
       \textbf{mathematics} & \textbf{condensed matter} & \textbf{neuroscience} & \textbf{astrophysics} & \textbf{quantum information}\\ \hline
        3.73\% & 3.48\% & 3.48\% & 3.23\% & 3.23\%\\ \hline \hline
  \end{tabular}
\end{table*}

\iffalse
\begin{figure}[!ht]
	\centering
	\subfloat[Hep-th]{\label{fig_m2} \includegraphics[width=0.22\textwidth]{img/HepthEdgeLabel_nc_d_128_macro}}
	\hfil 
% 	\subfloat[Twitter (Placeholder)]{\label{fig_m2} \includegraphics[width=0.32\textwidth]{img/HepthEdgeLabel_nc_d_128_macro}}
% 	\hfil
	\subfloat[20-Newsgroup]{\label{fig_m2} \includegraphics[width=0.22\textwidth]{img/20NewsEdge_nc_d_128_macro}}
	\hfil
    \subfloat[Hep-th]{\label{fig_m2} \includegraphics[width=0.22\textwidth]{img/HepthEdgeLabel_nc_d_128_micro}}
	\hfil
% 	\subfloat[Twitter (Placeholder)]{\label{fig_m2} \includegraphics[width=0.32\textwidth]{img/HepthEdgeLabel_nc_d_128_micro}}
% 	\hfil
	\subfloat[20-Newsgroup]{\label{fig_m2} \includegraphics[width=0.22\textwidth]{img/20NewsEdge_nc_d_128_micro}}
	\hfil
	\caption{Node classification results for different data sets.}
	\label{fig:nc}
    \vspace{-.5cm}
\end{figure}
\fi

Node classification refers to the task of predicting missing labels for nodes.
In these experiments, we evaluate the performance of our model on a multi-label node classification task, in which each node can be assigned one or more labels.
To test our model against the baselines, we use the node embeddings learned by the models as input to a one-vs-rest logistic regression using the LIBLINEAR library.
We vary the train to test ratio from 20\% to 90\% and report results on $micro-F1$ and $macro-F1$.
As for link prediction, we perform each split 5 times and plot mean and standard deviation.

We obtain the node labels for Hep-th collaboration network by searching for author interests using the Google Scholar API.
In the interest of keeping label dimensionality and class imbalance low, we consider the top 20 most common interests.
Table \ref{tab:hepth_labels} enumerates the top interests and percentage of authors with those interests.

Figure \ref{fig:nc} illustrates the results of our experiments.
For Hep-th, we observe that our model achieves higher $macro-F1$ than the baselines although it doesn't show much improvement over SDNE in the $micro-F1$ scores.
This can possibly be explained by the generality of most frequent labels.
As Hep-th is a theoretical physics conference, having interests such as ``theoretical physics'' are not surprising.
Better $macro-F1$ performance shows that our model can predict low occurring class labels better by utilizing the topics discussed in the abstract.
For 20-Newsgroup, our model outperforms baselines in both $micro-F1$ and $macro-F1$.

% \subsection{Visualization}

\subsection{Effect of Each Component}

As our model composes of several modules, we test the effect each component has on the prediction.
For this experiment, to test each module, we set the coefficients of the other components as zero.

% \begin{figure}[!htbp]
% \includegraphics[width=0.32\textwidth]{img/HepthEdgeLabel_component_lp_d_128_map}
%  \vspace{-.5cm}
% \caption{Effect of each component on MAP of link prediction for Hep-th. AE, VAE, HO and R stand for Auto Encoder, Variational Auto Encoder, Higher Order Proximity and Role based features respectively.}
% \label{fig:compEffect}
% % \vspace{-.5cm}
% \end{figure}

\iffalse
\begin{table}[!htbp]
	\centering\footnotesize
    \renewcommand{\arraystretch}{1.3}
    \caption{Effect of each component on link prediction for Hep-th.}
    \label{tab:compEffect}
    \vspace{-10pt}
  \begin{tabular}{ l | c  }
  %\begin{tabular}{| p{1.9cm} | l | p{1.5cm} | p{1.5cm} |}
           \hline \hline
      Algorithm & MAP\\ \hline \hline
      Autoencoder (AE) & 15\%\\ \hline
      Variational Autoencoder (VAE) & 15.2\%\\ \hline \hline
		\multicolumn{2}{l}{\emph{Adding features to VAE}} \\ \hline \hline
      + Higher Order (HO) & 21.6\%\\ \hline
      + HO + Roles (HO-R)& 22.2\%\\ \hline
      + HO-R + Node aggregated edge attributes (NA-ELAINE) & 22.7\%\\ \hline
      + HO-R + Edge attributes (ELAINE) & 23\%\\ \hline \hline
  \end{tabular}
\end{table}
\fi

\begin{table}[!htbp]
	\centering\footnotesize
    \renewcommand{\arraystretch}{1.3}
    \caption{Effect of each component on link prediction for Hep-th.}
    \label{tab:compEffect}
    \vspace{-10pt}
  \begin{tabular}{ l | c  }
  %\begin{tabular}{| p{1.9cm} | l | p{1.5cm} | p{1.5cm} |}
           \hline \hline
      Algorithm & MAP\\ \hline \hline
      Autoencoder (AE) & 15\%\\ \hline
      Variational Autoencoder (VAE) & 15.2\%\\ \hline 
      VAE+ Higher Order (HO) & 21.6\%\\ \hline
      VAE+ HO + Roles (HO-R)& 22.2\%\\ \hline
      VAE+ HO-R + Node aggr. edge attributes (NA-ELAINE) & 22.7\%\\ \hline
      VAE + HO-R + Edge attributes (ELAINE) & 23\%\\ \hline \hline
  \end{tabular}
\end{table}

Table \ref{tab:compEffect} illustrates the results.
We see that addition of Variational Autoencoder improves the MAP value by 0.2\% showing that effective regularization can positvely impact generalizability.
Adding higher order information benefits the most, increasing the MAP by 6.4 \%, followed by edge attributes and role based features which further improve MAP by 0.8\% and 0.6\% respectively.

We also compare our model against NA-ELAINE (Node Aggregated ELAINE) to show the loss the of information in aggregating edge labels for each node.
We observe NA-ELAINE improves MAP by 0.5\% whereas ELAINE achieves a higher value of 23\% achieving an increase of 0.8\% over the edge attribute unaware features illustrating the above effect.

\subsection{Hyperparameter Sensitivity}
\begin{figure}
    \centering
    \includegraphics[width=0.32\textwidth]{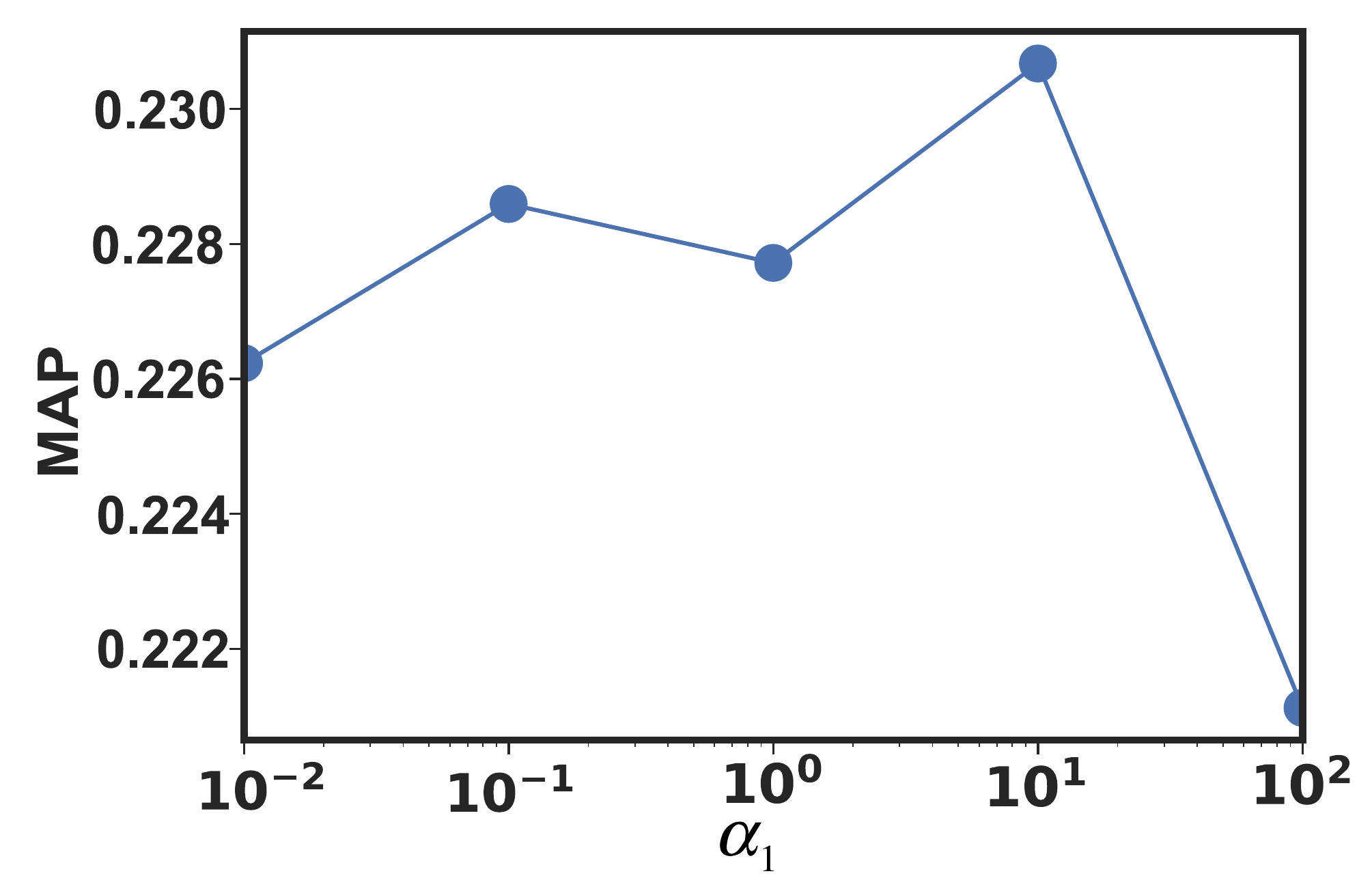}
     \vspace{-.5cm}
    \caption{Effect of $\alpha_1$, coefficient of edge label reconstruction, on link prediction MAP.}
    \label{fig:hyp_alpha_1}
\end{figure}
\begin{figure}
    \centering
    \includegraphics[width=0.31\textwidth]{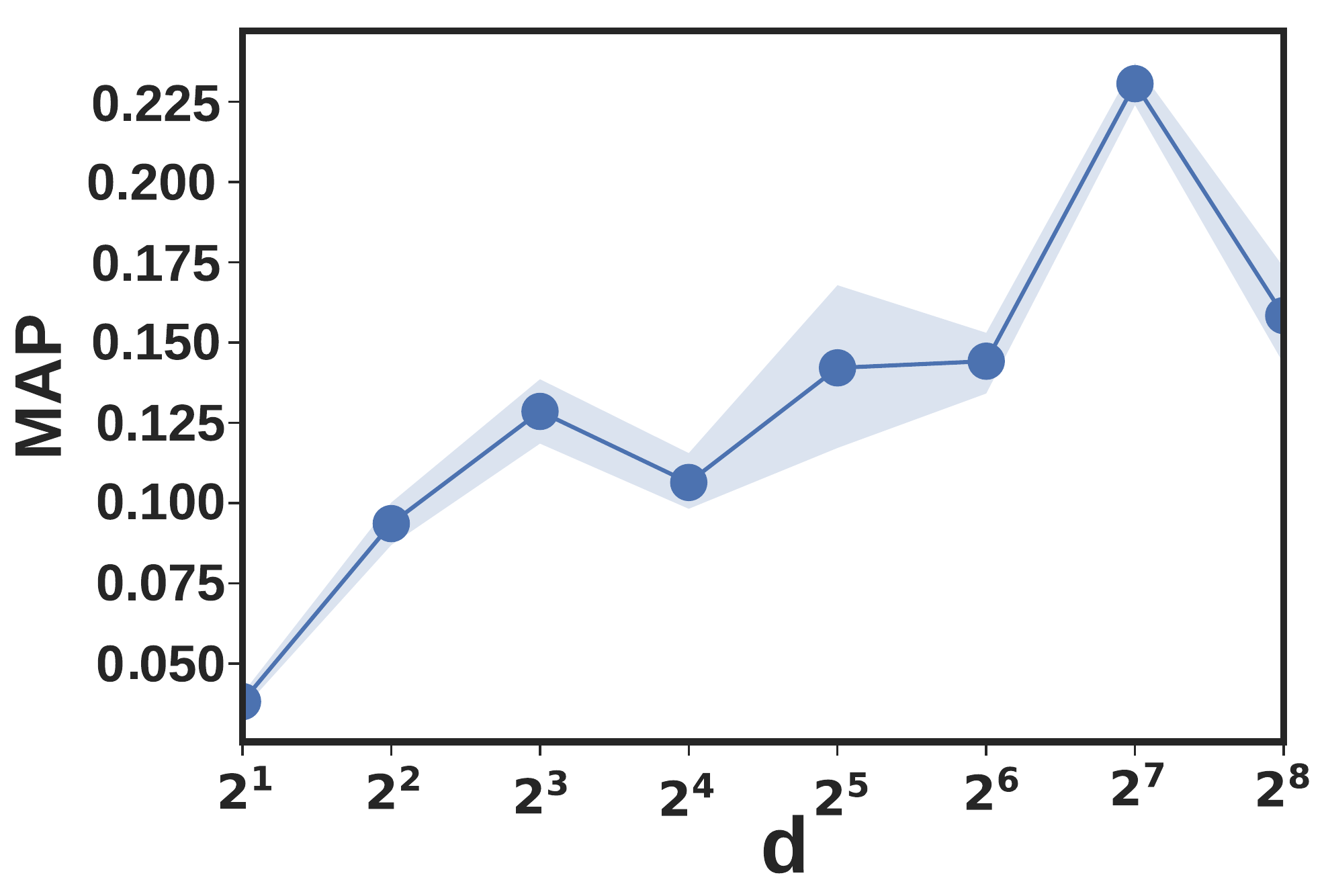}
     \vspace{-.5cm}
    \caption{Effect of embedding dimensions on link prediction MAP. It shows that link prediction performance peaks at 128.}
    \label{fig:hyp_embed_dim}
\end{figure}
In this set of experiments, we evaluate the effect of hyperparameters on the performance to understand their roles.
Specifically, we evaluate the performance gain as we vary the number of embedding dimensions, $d$, and the coefficient of edge label reconstruction, $\alpha_1$.
We report MAP of link prediction on Hep-th for these experiments.

\textbf{Effect of dimensions}: We vary the number of dimensions from 2 to 256 in powers of two. 
Figure \ref{fig:hyp_embed_dim} illustrates its effect on MAP.
As the number of dimensions increases the link prediction performance improves until 128 as higher dimensions are capable of storing more information.
The performance degrades as we increase the dimensions further as the model overfits on the observed edges and performs poorly on predicting new edges.

\textbf{Effect of $\alpha_1$}: The value of $\alpha_1$ determines the balance between neighborhood prediction and edge attribute prediction.
We test the values from $10^{-2}$, which prioritizes neighborhood and edge attribute, to $10^2$, which penalizes edge label reconstruction loss more heavily.
Figure \ref{fig:hyp_alpha_1} shows the relation of MAP with $\alpha_1$.
We observe that initially as we increase $\alpha_1$, MAP increases which suggests that the model benefits from having link attributes.
Increasing it further by 10 times drastically reduces the performance as now the embedding almost solely represents the edge labels which on their own cannot predict missing labels well.
This demonstrates that having edge labels can significantly improve link prediction performance.

%% file: src/discussion.tex
\section{Conclusion}\label{sec:discussion}
In this paper, we presented ELAINE, a higher-order proximity and social role preserving network embedding which utilizes edge attributes to learn a unified representation.
It jointly optimizes the reconstruction loss of neighborhood, role and edge attribute based features.
It uses a coupled deep variational autoencoder to capture highly non-linear nature of interactions between nodes and an edge attribute decoder to reconstruct the edge labels.
Our experiments demonstrate the efficacy of our model over several real world data sets including collaboration networks and social networks on challenging tasks of link prediction and node classification.

There are several promising directions for future work: (1) predicting missing edge labels, (2) extending the model for dynamic graphs, (3) automatic optimization of hyperparameters, and (4) better regularization.
As our work jointly models the edge attributes and network structure, it can be extended to predict the edge attributes over the missing edges. The model also can be extended to represent dynamic networks in vector space~\cite{dai2017deep,goyal2017dyngem,zhu2016scalable} which can be useful for tasks like anomaly detection.
Also, given the number of hyperparameters, automatically choosing the optimal values can be of great significance.
As we illustrate that variational autoencoder can improve the performance, we would like to explore what effect different regularizations can have on the performance.